# Exactly solvable model of a quantum spin glass

Th. M. Nieuwenhuizen

*Van der Waals-Zeeman Laboratorium, Universiteit van Amsterdam*
*Valckenierstraat 65, 1018 XE Amsterdam, The Netherlands*

(Revised December, 1994)

A mean field spherical model with random couplings between pairs, quartets, and possibly higher multiplets of spins is considered. It has the same critical behavior as the Sherrington-Kirkpatrick model. It thus exhibits replica symmetry breaking. The order parameter function is solved exactly in the whole low temperature phase. The zero field cooled susceptibility remains finite at low $T$. Next a quantum version of the system is considered. Whereas the magnetic properties are not altered qualitatively, the thermodynamics is now regular at small temperatures.

7510.Nr, 7510.Jm, 7540.Cx, 7550.Lk

The phenomenon of spin glasses has posed a major new field of research. The basic phenomenon is breaking of ergodicity. This leads to highly non-trivial properties as observed recently e.g. in aging experiments [1] The most known spin glass model was formulated by Sherrington and Kirkpatrick (SK) [2]. Its solution problem was presented by Parisi, see [3] for a review. It gave a description of ergodicity breaking in terms of overlaps. For more recent reviews on spin glasses, see [4], [5], [6].

The spherical model has always played a special role for understanding the basic phenomena of phase transitions. It was introduced by the late Marc Kac and solved by Berlin and Kac. [7] It constituted a simple model for studying critical behavior, see Joyce for a review. [8].

The spherical random bond spin glass was studied in ref [9]. It shows freezing but no replica symmetry breaking. This model was extended by the present author to include short range ferromagnetic interactions. [10] The spherical model with random couplings between $p$ spins was studied recently by Crisanti and Sommers (CS). For $p > 2$ it was found that there occurs a one-step replica symmetry breaking. [11] Dynamical aspects of this transition were studied in detail. [12]

For Ising spin glasses the Parisi solution is known explicitly near the spin glass transition. In the frozen phase the solution has been put in the form of a stochastic differential equation, for which no explicit solution is known. As a model solvable in the whole low temperature phase we propose here a mean field spherical model with random couplings between pairs, quartets, and, possibly, higher multiplets of spins. In contrast to previous spherical models, where only one type of these couplings occurs, the model belongs to the universality class of the Ising spin glass. We shall derive the explicit form of the order parameter in the low temperature phase. Then we shall consider the recent quantum formulation of the spherical model [13] and study its low temperature behavior.

Consider a system with $N$ spins with random couplings between sets of $p$ spins. The Hamiltonian reads

$$\mathcal{H} = -\sum_{p=2}^{\infty} \sum_{i_1 < i_2 < \cdots < i_p} J_{i_1 i_2 \cdots i_p} S_{i_1} S_{i_2} \cdots S_{i_p} - H \sum_i S_i \qquad (1)$$

The $J$'s are independent Gaussian random variables with average zero and variances $\langle J^2_{i_1 i_2 \cdots i_p} \rangle = (p-1)! J_p^2 N^{1-p}$. The spins are subject to the spherical constraint $\sum_i S_i^2 = N\sigma$. The *classical* partition sum reads

$$Z_{cl} = \int DS \, e^{-\beta \mathcal{H}} \qquad (2)$$

where $DS = \delta(\sum_{i=1}^N S_i^2 - N\sigma) \prod_i (dS_i/\sqrt{\pi})$. The thermodynamics of this model with only one of the $p$ terms was worked out by CS. In a replica formulation of the average partition sum it was found that the free energy only depends on the overlap $q_{\alpha\beta} = (1/N) \sum_{i=1}^N S_i^\alpha S_i^\beta$. This result has to be summed over $p$. We introduce

$$f(q) = \sum_{p=2}^{\infty} \frac{1}{p} J_p^2 q^p \qquad (3)$$

and we obtain a replicated free energy

$$2\beta F_n = -\beta^2 \sum_{\alpha,\beta=1}^n [f(q_{\alpha\beta}) + H^2 q_{\alpha\beta}] - \sum_{\alpha=1}^n \{\log q\}_{\alpha\alpha} - n \qquad (4)$$

If $J_2 = 0$ the model has the one-step replica symmetry breaking solution studied by CS. When $J_2 > J_2^*(J_3, J_4, \cdots)$ is large enough, a continuous transition will take place. From now on we assume that this is the case. In eq. (4) also the diagonal terms $q_{\alpha\alpha} \equiv q_d$ occur; in the classical situation they are equal to $q_d = \sigma$. Expanding in off-diagonal elements $q_{\alpha\beta}$ one obtains

$$2\beta F_n = -n\{1 + \log q_d + \beta^2 f(q_d) + \beta^2 H^2 q_d\} - \beta^2 H^2 \sum{}' q_{\alpha\beta} - \frac{1}{2}(\beta^2 J_2^2 - \frac{1}{q_d^2}) \sum{}' q_{\alpha\beta}^2$$
$$- \frac{1}{3 q_d^3} \sum_{\alpha\beta\gamma}{}' q_{\alpha\beta} q_{\beta\gamma} q_{\gamma\alpha} - \frac{1}{3}\beta^2 J_3^2 \sum{}' q_{\alpha\beta}^3 - \frac{1}{4}\beta^2 J_4^2 \sum{}' q_{\alpha\beta}^4 + \cdots \tag{5}$$

where the primes indicate exclusion of diagonal terms. In the case $J_3 = 0$, but $J_4 > 0$, this is exactly the relevant part of the free energy functional of the SK model [14] [15]. The model thus belongs to the same universality class, and exhibits replica symmetry breaking. This continuous transition sets in at $T_G = J_2 q_d = J_2 \sigma$. For $J_3 > 0$ the model belongs to the universality class of the random bond Potts model. When the $J_2 > J_2^*$ and the next non-zero term is $J_p$ for some $p > 4$, the model belongs to a new universality class. We shall not consider such situations here and assume $J_2 > J_2^*$, $J_3 = 0$, $J_4 > 0$. The higher couplings are only relevant well below $T_G$.

We express the $q_{\alpha\beta}$ in the Parisi function $q(x)$. It takes the plateau value $q_1$ for $x_1 < x < 1$. In an external field it also has a plateau value $q(x) = q_0$ for $0 < x < x_0$. The inverse function is $x(q)$. Using the expressions of an appendix of ref. [11] we obtain the explicit expression for the *classical* free energy

$$2\beta F_{cl} = -\beta^2 \int_0^1 dx \{f(q_d) - f(q(x)) + H^2 q_d - H^2 q(x)\} - \int_0^{q_1} \frac{dq}{I(q)} - \log(q_d - q_1) - 1 \tag{6}$$

where

$$I(q) = q_d - q_1 + \int_q^{q_1} x(q') dq' \tag{7}$$

The saddle point equation for $q(x)$ reads

$$\beta^2 f'(q(x)) + \beta^2 H^2 = \int_0^{q(x)} dq' \frac{1}{I(q')^2} \tag{8}$$

In the region where $q'(x) \neq 0$ one has $\beta^2 f''(q) = I(q)^{-2}$. It follows that $x(q)$ has a universal shape at all $T$

$$x(q) = T \frac{f'''(q)}{2\{f''(q)\}^{3/2}} \tag{9}$$

which gives $q(x; T) = q(\beta x)$ after inversion. In an external field $H$ the plateau value $q_0$ follows from

$$H^2 = q_0 f''(q_0) - f'(q_0) \tag{10}$$

This condition is independent of $T$. $q_1$ follows from

$$q_1 = q_d - \frac{T}{\sqrt{f''(q_1)}} \tag{11}$$

Of special interest is the case where the couplings $J_p$ are such that

$$f(q) = J_2^2 \frac{q}{4a} \log \frac{1+aq}{1-aq} \tag{12}$$

for some $a$ in the range $0 < a < 1/\sigma$. In zero field one finds $q(x) = (\beta J_2/2a^2)x$ for $0 < x < x_1$ and

$$q_1 = \frac{2(q_d J_2 - T)}{J_2 + \sqrt{J_2^2 - 4a^2 T(q_d J_2 - T)}} \tag{13}$$

Note that here the order parameter function $q(x)$ is either linear or constant *in the whole low temperature phase*. The breakpoint $x_1 = 2a^2 T q_1$ vanishes both near the critical temperature and near zero temperature.

In an external field the freezing transition is given by the AT-line [16]. Here the *replicon* mode [15] becomes massless. We propose to give this mode the more physical name *ergodon*. The critical field where this occurs, $H_{fr}(T)$, is obtained by inserting $q_0 = q_1(T)$ into eq. (10). Note that $H_{fr}$ remains finite at $T = 0$.

In the high temperature phase the susceptibility follows the Curie law $\chi = \beta q_d$. The zero field cooled susceptibility reads

$$\chi_{ZFC} = \beta(q_d - q_1) = \frac{1}{\sqrt{f''(q_1(T))}} \qquad (14)$$

It remains finite at $T = 0$. This is usually observed experimentally. However, the SK model predicts a vanishing value at $T = 0$. [17] The field cooled susceptibility

$$\chi_{FC} = \beta \int_0^1 dx(q_d - q(x)) = \frac{1}{J_2} \qquad (15)$$

is constant in the frozen phase, in good agreement with experiments. The internal energy and the entropy read

$$U_{cl} = \beta[f(q_1) - f(q_d)] + \frac{f'(q_1)}{\sqrt{f''(q_1)}} - \int_0^{q_1} dq\sqrt{f''(q)} \qquad (16)$$

$$2\,S_{cl} = \beta^2[f(q_1) - f(q_d)] + \frac{\beta f'(q_1)}{\sqrt{f''(q_1)}} + 1 + \log(q_d - q_1)$$

For small $T$ one has $S_{cl} \approx \frac{1}{4} - \frac{1}{4}\log\beta^2 f''(q_1)$. As usual for classical vector or spherical spins, it goes to $-\infty$ as $T \to 0$. This also implies that the specific heat goes to a constant, $C \to 1/2$, at low $T$. For the case $\sigma = 1/2$ with pair and quartet couplings ($J_2 = 1$, $J_4 = 2$), the specific heat and the entropy have been plotted in Figure 1.

We have wondered whether the anomalous low temperature behavior of our model can be cured. In order to do so, one should regularize the low temperature behavior of the spherical model in general. This can be done by going to a description of spherical spins in a standard thermal field theory for bosonic *quantum* spins. [13] The situation becomes simplest when the spins are complex valued and couplings are Hermitean. For the case of random pair and quartet couplings we now assume the Hamiltonian

$$\mathcal{H} = -\frac{1}{2}\sum_{i,j} J_{ij} S_i^* S_j - \frac{1}{24}\sum_{i,j,k,l} J_{ijkl} S_i^* S_j S_k^* S_l - H\sum_i (\text{Re}S_i + \text{Im}S_i) \qquad (17)$$

For each pair $(i,j)$ there are now two independent random variables, $J'$ and $J''$, with average zero and variance $J_2^2/2N$, in terms of which $J_{ij} = J_{ji}^* = J' + iJ''$. For each quartet $(i,j,k,l)$ with $i < k$, $j < l$ there are four independent random variables, $J'_{1,2}$ and $J''_{1,2}$, each having average zero and variance $9J_4^2/2N^3$. In terms of $J_{1,2} = J'_{1,2} + iJ''_{1,2}$ the couplings in eq. (17) read $J_{ijkl} = J_{jilk}^* = J_1 + iJ_2$ and $J_{ijlk} = J_{jikl}^* = J_1 - iJ_2$. The thermal partition sum involves spins on the imaginary-time interval $0 < \tau < \beta$

$$Z = \int \mathcal{D}S \exp\int_0^\beta d\tau\left\{\frac{-1}{4\alpha}\sum_{i=1}^N S_i^*(\tau)\frac{d}{d\tau}S_i(\tau) - \mathcal{H}(S_i(\tau))\right\} \qquad (18)$$

with boundary conditions $S_i(\beta) = S_i(0)$. The $\tau$-axis is discretized in $\mathcal{M} = \beta/d\tau$ steps, with first $N \to \infty$ and then $\mathcal{M} \to \infty$, while $dS(\tau) = S(\tau + d\tau) - S(\tau)$. The integration measure is a repetition of the spherical measure at all $\tau$, $\mathcal{D}S = C^N\prod_\tau DS(\tau)$, where $DS = \delta(2N\sigma - \sum_i S_i^* S_i)\Pi_i(dS_i^* dS_i/2\pi)$ is the spherical measure for complex spins and $C = (1/2\alpha)\Pi_{n\neq 0}(\pi|n|/2\alpha)$.

For a ferromagnet eq. (18) leads to similar critical behavior as the classical spherical model. However, the entropy is non-negative and vanishes at $T = 0$. The zero point magnetization, $M_0 = \sqrt{\sigma - \alpha}$, shows a quantum reduction from the classical value $M_0 = \sqrt{\sigma}$. [13]

Here we wish to see whether the low temperature behavior is also cured for our spin glass model. We therefore extend the Crisanti-Sommers approach to thermal fields. We introduce a "Fermi level" $\mu_\alpha(\tau)$ related to the spherical constraint at imaginary time $\tau$ and overlaps $q_{\alpha\beta}(\tau,\tau') = (1/N)\sum_i S_i^*(\tau)S_i(\tau')$. We look for time-invariant solutions where $\mu_\alpha(\tau) = \mu$ and where $q_{\alpha\beta}(\tau,\tau') = \sum_\epsilon q_{\alpha\beta}(\epsilon)\exp\{2\pi inT(\tau - \tau')\}$ with bosonic Matsubara frequencies $\epsilon \equiv \pi nT/2\alpha$. In the high temperature phase at zero field only the $q_{\alpha\alpha}(\epsilon)$ are non-zero. We shall denote $q_d \equiv q_{\alpha\alpha}(\epsilon = 0)$, which now becomes a smooth, increasing, strictly positive function of temperature. The spin glass temperature $T_G$ is the solution of the relation $T_G = J_2 q_d(T_G)$. Here the off-diagonal elements $q_{\alpha\beta}(\epsilon = 0)$ become non-zero; since $T_G < \sigma$, this occurs below the classical transition point. The $q_{\alpha\beta}(\epsilon)$ with $\epsilon \neq 0$ are always equal to zero. [18] For $\epsilon \neq 0$ we set $p_\epsilon \equiv \beta q_{\alpha\alpha}(\epsilon)$ and find that the full free energy reads

$$F = 2F_{cl}(q(x); q_d) + F_{qc}(q_d; \mu; p_\epsilon) \tag{19}$$

where $F_{cl}$ is the free energy given by eq. (6) with $f(q) = J_2^2 q^2/2 + J_4^2 q^4/4$. (The factor 2 in eq. (19) arises since now spins are complex.) The quantum correction equals

$$\beta F_{qc} = -\sum_{\epsilon \neq 0}\left\{\frac{J_2^2}{2}p_\epsilon^2 + 1 - (\mu + 2i\epsilon)p_\epsilon + \log 2i\epsilon p_\epsilon\right\} - \frac{T^2 J_4^2}{4}\sum_{\epsilon_1+\epsilon_2=\epsilon_3+\epsilon_4} p_{\epsilon_1} p_{\epsilon_2} p_{\epsilon_3} p_{\epsilon_4} + \beta\mu(q_d - \sigma) + \log 2\alpha \tag{20}$$

where terms with all $\epsilon_i$ equal to zero are excluded; they occur already in $F_{cl}$. In the quantum situation $q_d$, $\mu$, and all $p_\epsilon$ have to be treated as additional variational parameters. For large temperatures one finds $\mu = Ty/\alpha$, $q_d = \alpha/y$, $p_\epsilon = 1/(\mu + 2i\epsilon)$, with $\tanh y = \alpha/\sigma$. The entropy $S_\infty = y\sigma/\alpha - \log 2\sinh y$ agrees with the result in the situation without interactions. [13] Only in the classical limit $\alpha \to 0$ one finds back previous relation $q_d = \sigma$. Therefore the susceptibility $\chi = q_d(T)/T$ has been reduced from its classical value at any $T$.

First consider the situation where only pair couplings occur, $J_4 = 0$. This case, where no replica symmetry breaking occurs, is exactly solvable since only Gaussian integrals occur. The free energy may be expressed as

$$\beta F = -\beta\mu\sigma + \int d\lambda \rho(\lambda) \log 2\sinh \alpha\beta(\mu - \lambda) \tag{21}$$

Here $\rho(\lambda)$ denotes the semi-circular density of eigenvalues of the coupling matrix $J_{ij}$, viz. $\rho(\lambda) = \sqrt{4J_2^2 - \lambda^2}/2\pi J_2^2$. Alternatively we can solve the $p_\epsilon$ from eq. (20). They read $p_\epsilon = 1/[\frac{1}{2}\mu + i\epsilon + \sqrt{(\frac{1}{2}\mu + i\epsilon)^2 - J_2^2}]$ with $\mu = 2J_2$ in the condensed phase. When expanding the log sinh in eq. (21) as an $\epsilon$-sum, we find a term-by-term agreement with eq. (20). It is clear that $S_0 = 0$, showing that our quantum formulation indeed regularizes the low $T$ anomaly of the entropy of ref. [9]. The entropy and specific heat now behave as $S \sim C \sim T^{3/2}$, related to the square root singularity of the semi-circular law at $\lambda = 2J_2$. This describe gapless excitations. In contrast, the mean field quantum spherical ferromagnet does have a gap. [13]

When also quartet couplings, and possibly higher couplings, are present, replica symmetry is broken. The entropy at low $T$ can be obtained by analyzing the continuum limit of the equations for $p_\epsilon$. Details of this derivation will not be presented here. It turns out that the same behavior occurs, $S \sim T^{3/2}$, $C \sim T^{3/2}$.

Figure 2 presents the specific heat and entropy for $\alpha = 1/4$ in the situation of Fig. 1, $\sigma = 1/2$, $J_2 = 1$, $J_4 = 2$.

In conclusion, we have presented a set of models where the order parameter function can be solved exactly in the whole low temperature phase. The critical behavior is exactly the one derived by Parisi for the SK model. As compared to that model, the solution is explicit for all $T$. It is shown that the order parameter function has a universal shape in the whole low temperature region. The breakpoint $x_1$ is small near the transition and near $T = 0$. Well below $T_G$ states with very different overlaps, so of very different nature, occur.

We have also considered a quantum version of the model. It is seen that the shape of the order parameter function is not changed. The low $T$ divergency of the entropy in the classical model is eliminated. Now the entropy and specific heat vanish as $T^{3/2}$.

It is interesting to point out that the magnetic properties of the model, such as $q(x)$, $\chi_{ZFC}$ and $\chi_{FC}$ have not changed qualitatively in the quantum description. The only relevant difference is that the self-overlap $q_d$ has become a smooth function of $T$. In our model $\chi_{ZFC}$ remains non-zero at $T = 0$. This disagrees with the SK model but is very often observed experimentally. Another new aspect of our model is the boundedness of the critical field below which freezing takes place. For non-zero field one has in the frozen phase $M(H, T) = H/\sqrt{f''(q_0)} = M(H)$ and $S(T, H) = S(T)$. Therefore the Parisi-Toulouse hypothesis is satisfied. [19]

It is hoped that the proposed model can be used as a basis to explain experiments. Hereto inclusion of ferromagnetic couplings and extension to vector spins seems of large interest. In a dynamical study of the model a closed set of equations will occur that bear the same physics as the more complicated set of the SK-model.

## ACKNOWLEDGMENTS

The author is grateful for discussions with J.H. Sommers, H. Horner, and J.A. Mydosh. He thanks J.M. Luck and N. Garcia for hospitality at the CEA, Saclay, and the Universidad Autonoma de Madrid, where part of this work was done. This work was made possible by the Royal Dutch Academy of Arts and Sciences (KNAW).

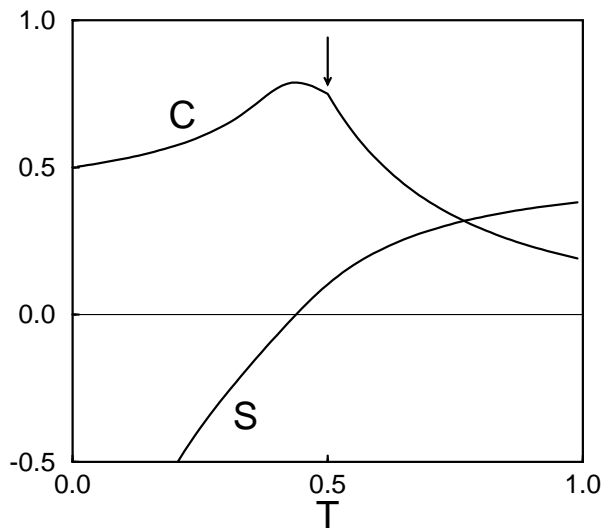

FIG. 1. Specific heat and entropy as function of temperature in the classical model. The arrow marks $T_G$.

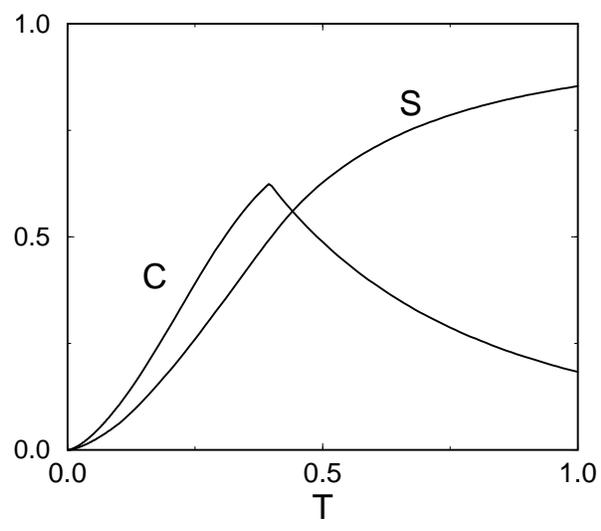

FIG. 2. Specific heat and entropy in the quantum model.